\documentstyle[epsfig,macros]{prhep97}

\makeatletter
\let\chapter\hid@chapter
\makeatother
\begin{document}
%%\pagenumbering{empty}

% The following definitions need to be customized;

% Will appear on page headings
\authorrunning{R.\,Sommer}
\titlerunning{ The $\Lambda$-parameter and $m_s$ of quenched QCD}
 
% Now the full name of author and talk

% For plenary talks, the talk number is that of the session
\def\talknumber{1507} 

\title{The $\Lambda$-parameter and $m_s$ of quenched QCD$^1$}
\author{Rainer\,Sommer 
(sommer@ifh.de) for the ALPHA collaboration}
\institute{ DESY-IfH, Platanenallee 6, D-15738 Zeuthen, Germany}

\maketitle
\vspace{-4.5cm} 

{ \normalsize
\hfill \parbox{21mm}{DESY 97-223}}\\[32mm]

\addtocounter{footnote}{1}
\footnotetext{
Talk presented at the {\it International Europhysics Conference
on High Energy Physics}, Jerusalem, August 19--26 1997.
}

\begin{abstract}
We explain how scale dependent renormalized quantities 
can be computed using lattice QCD. Two examples are used: the running 
coupling and quark masses. A reliable computation of the 
$\Lambda$-parameter in the quenched approximation is presented.
\end{abstract}
\section{Introduction}
In a perturbative treatment, one takes as parameters 
of QCD the  running coupling and running quark masses in a specific 
renormalization scheme. If we adopt a mass-independent scheme,
the coupling and quark masses satisfy rather simple renormalization group 
equations. Their asymptotic high energy behavior is given by 
the $\Lambda$-parameter and the renormalization group invariant masses 
$\{M_f\}$ 
($f$ labeling the flavors).
These parameters have the advantage that 
$M_f$ are independent of the renormalization scheme and 
the $\Lambda$-parameters of different schemes can be related exactly 
by 1-loop calculations. 

Once one considers QCD on the non-perturbative level, 
it is natural to take $\nf+1$ hadronic observables to be the
parameters that fix the theory instead of $\Lambda$ and $\{M_f\}$. We call
this a hadronic scheme (HS).
For QCD with $u$ and $d$ taken as
mass-degenerate and an additional $s$-quark, 
one may take for example the proton mass, $m_{\rm p}$, the pion mass, $m_\pi$,
and the Kaon mass, $m_{\rm K}$, as the basic parameters. 
%% In fact, this is
%% also a choice that is of practical advantages, when one wants to
%% make predictions for QCD through Monte Carlo simulations of
%% the lattice regularized path integral. 
Of course, the two sets of parameters
are related once a non-perturbative solution of QCD is available.
It is a challenge for lattice QCD to provide this relation with
completely controlled errors. In this brief report we describe a 
method that allows to solve this problem. 
The numerical results (sect. 4) are for quenched QCD, i.e. the fermion
 determinant is omitted
in the path integral; one may think of this as a version
of QCD with $\nf=0$ flavors of sea-quarks.

The problem described above is the
connection of the short distance and long distance dynamics of QCD.
As such it appears at first sight to be intractable by numerical simulations,
since it requires that { various scales} are treated 
{ on one and the same lattice}. 
This is { impossible}, if discretization errors are supposed to be 
under control!
However, one may define a running coupling and running quark masses
in QCD in a finite volume of linear size $L$ with suitably chosen
boundary conditions (cf. sect. 3). 
Then the masses and coupling run with an energy
scale $q=1/L$ and their {\it evolution can be computed recursively}. 
The only
requirement to keep discretization errors small 
in each step of the recursion is $L \gg a$ which is easy to
satisfy ($a$ denotes the lattice spacing)\cite{alphaI,alphaIII}.
\section{Strategy}
\begin{figure}[ht]
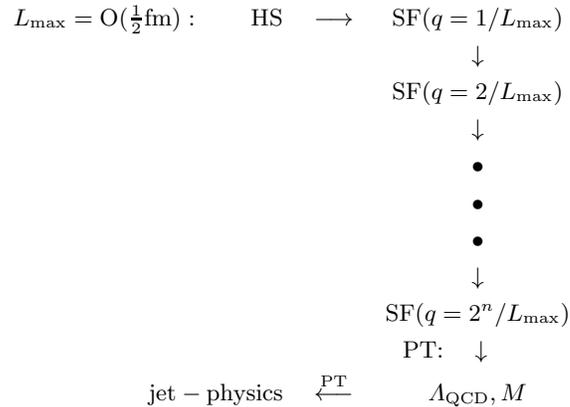

\vspace{-0.5cm}
\bes
 { L_{\rm max}}=\rmO(\frac{1}{2}\fm): \qquad 
 {\rm HS} \quad \longrightarrow \quad
      &{\rm SF} (q=1/{ L_{\rm max}})& \quad 
               \nonumber \\
      &\downarrow&  \nonumber \\
      &{\rm SF} (q=2/{ L_{\rm max}})&  \nonumber \\ 
      &\downarrow&  \nonumber \\
      &\bullet&  \nonumber \\
      &\bullet&  \nonumber \\
      &\bullet&  \nonumber \\
      &\downarrow&  \nonumber \\
      &{\rm SF} (q=2^n/{ L_{\rm max}})& \nonumber \\
   &\mbox{ \small PT:} \quad  \downarrow \qquad \quad &  \nonumber \\
{\rm jet-physics} \quad \stackrel{\rm  PT}{\longleftarrow} 
     \quad   &\Lambda_{\rm QCD}, M & \nonumber
\ees
\vspace{-0.8cm}
\caption{The strategy for a non-perturbative computation of 
         short distance parameters.
\label{f_strategy}}
\end{figure}

We start by giving an overview of 
the strategy to compute short distance parameters 
in \fig{f_strategy}.
%%%%%%%%%%%%%%%%%%%%%%%%%%%%%FIGURE%%%%%%%%%%%%%%%%%%%%%%%%%%%%%%%%%%%
%%%%%%%%%%%%%%%%%%%%%%%%%%%%%%%%%%%%%%%%%%%%%%%%%%%%%%%%%%%%%%%%%%%%
One first renormalizes QCD replacing the bare parameters by hadronic
observables. This defines the HS introduced above.
It can be
related to the finite volume scheme (denoted by SF)
at a low energy scale $q=1/L_{\rm max}$, where $L_{\rm max}$ is
of the order of  $1~\fm$. 
Within this scheme one then computes the scale evolution
up to a desired energy $q=2^n/{ L_{\rm max}}$. It 
is no problem to choose the number of steps $n$ large enough to be
sure that one has entered the perturbative regime. There 
perturbation theory (PT) is used to evolve further to infinite energy and
compute $\Lambda$  and $\{M_f\}$.
Inserted into perturbative expressions, these parameters provide
predictions for jet cross sections or other high energy observables. 
In \fig{f_strategy}, all
arrows correspond to relations in continuum QCD;
the whole strategy is designed such that lattice calculations
for these relations can be extrapolated to the continuum limit.  

For the practical success of the approach, the finite volume coupling
and quark masses must satisfy a number of criteria.\vspace{-0.1cm}
\begin{itemize}
     \item{They should have an easy perturbative expansion, such that
           the $\beta$-function (and $\tau$-function, which describes the 
           evolution of the running masses) can be computed to 
           sufficient
           order in the coupling.} 
     \item{They should be easy to calculate in MC (small variance!).}
     \item{Lattice effects must be small to allow
           for safe extrapolations $a/L \to 0$.}   
\end{itemize}\vspace{-0.1cm}
Careful consideration of the above points led to the introduction
of the running coupling, $\gbar$, and quark mass, $\mbar$, through the 
Schr\"odinger functional (SF) of QCD 
\myref{alphaII,alphaIII,StefanI,lat97}.
%We introduce this scheme in the following section. 

\section{Schr\"odinger functional scheme}
We describe the
definition of  $\gbar$ and $\mbar$
in a formal continuum formulation. The SF is the 
partition function of the Euclidean path integral on a hyper-cylinder
with Dirichlet boundary conditions
in time, ($P_{\pm}=\frac12(1\pm\dirac{0})$)
\bes 
\left.
 \begin{array}{llll}
  A_k(x)=  \bvalue_k({\bf x}),\quad& 
   P_+\psi(x) = \rho({\bf x}),\quad &
  \psibar(x) P_-= \rhobar({\bf x})\quad &
    \mbox{at} \, \, x_0=0  \enspace ,
     \\
  A_k(x)=  \bvalue'_k({\bf x}),\quad & 
   P_-\psi(x) = \rho'({\bf x}),\quad &
  \psibar(x) P_+= \rhobar'({\bf x})\quad & 
    \mbox{at} \, \, x_0=L  \enspace .
    \end{array}
    \right.
\label{e_bcA}
\ees
Here, $A_k, \ldots, \rhobar'$ are
classical prescribed boundary fields.
In space, the gauge fields are taken
periodic under shifts $x \to x+L \hat{k}$
and the fermion fields are 
periodic up to a phase $\theta$.

The renormalized coupling is defined through the response of the 
SF to an infinitesimal change 
of the boundary gauge fields $C_k,C_k'$. Since the boundary fields
are taken with a strength proportional to $1/L$, there is no scale apart from
$L$ and the coupling $\gbar(L)$ runs with $L$.
Details have been discussed in several publications \myref{alphaII,alphaIII}. 

A natural starting point 
for the definition of a renormalized quark mass
is the PCAC relation,
\bes
 \partial_\mu A_{\mu}^a(x) = 2 m P^a(x)\enspace ,
 \label{e_PCAC}
\ees
which expresses the proportionality of the divergence
of the axial current, $A_{\mu}^a$,
to the pseudo-scalar density $P^a$, where
\bes            
  P^a(x) = 
            \psibar(x)\dirac{5}\frac{1}{2}\tau^a\psi(x), \quad
  A_{\mu}^a  =
            \psibar(x)\dirac{\mu}\dirac{5}\frac{1}{2}\tau^a\psi(x) 
           \enspace ,
  \label{e_density}          
\ees
and $\tau^a$ are the Pauli-matrices acting in flavor space (we take
$\nf=2$ degenerate flavors from now on).
Renormalizing the operators in \eq{e_PCAC},
\bes
 (\ar)_{\mu}^a = \za A_{\mu}^a \enspace ,\quad
  \pr^a = \zp P^a \enspace , \label{e_zazp}
\ees
we define a renormalized quark mass by
\bes
 \mbar = {\za \over \zp } m \enspace . 
 \label{e_mbar}
\ees
Here, $m$ is to be taken from \eq{e_PCAC} inserted into an arbitrary 
correlation function. It does not depend on the chosen correlation
function, since \eq{e_PCAC} is an 
operator identity. The scale independent renormalization
$\za$ can be fixed through current algebra relations (also in the lattice
regularization \myref{Boch,paper4}) and we are left to give a 
normalization condition for the pseudo-scalar density, $\pr$.
The running mass, $\mbar$, then inherits its scheme-
and scale-dependence from the corresponding dependence of
$\pr$. 

We define $\zp$ in terms of
correlation functions in the SF.
%%%%%%%%%%%%%%%%%%%%%%%%%%%%%FIGURE%%%%%%%%%%%%%%%%%%%%%%%%%%%%%%%%%%%
\begin{figure}[ht]
\vspace{ -0.0cm}
\centerline{
\psfig{file=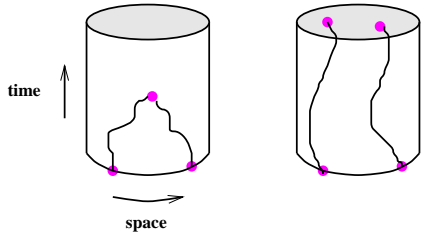,%
width=5.5cm}}
\vspace{-0.0cm}
\caption{$\fp$ (left) and $f_1$ (right) in terms of quark propagators.
\label{f_matrixelements}}
\end{figure}
%%%%%%%%%%%%%%%%%%%%%%%%%%%%%%%%%%%%%%%%%%%%%%%%%%%%%%%%%%%%%%%%%%%%
To start, we construct (isovector) pseudo-scalar fields at the 
boundary of the SF \footnote{
At this point,
the SF provides us with an important advantage compared to other 
boundary conditions, namely we can project the boundary quark fields 
onto zero momentum. As a result, the correlation functions vary slowly 
with $x_0$, leading to both small statistical and small
discretization errors.},
\bes
  \op{}^a &=& \int \rmd^3 {\bf u} \int \rmd^3 {\bf v}
  \,\,\zetabar({\bf u})\dirac{5}\frac{1}{2}
  \tau^a\zeta({\bf v})
  \enspace,  \label{e_boundops}
\ees
from the ``boundary quark fields'', $\zeta,\zetabar$. Their
precise definition \cite{paper1} involves a functional derivative,
e.g. $\zeta({\bf x})\equiv \frac{\delta}{\delta \rhobar({\bf x})}$. After 
taking this functional derivative one sets the boundary values, 
$\rho ...$ ,to zero.
Analogously one defines a field $ \op{}'^a$, which resides at $x_0=L$.
These fields are used in the correlation functions 
\bes
 \fp(x_0) = - \frac13 \langle P^a(x)  \op{}^a\rangle \enspace ,
 \quad
 f_1 &=& \langle\opprime{}^a\op{}^a\rangle \enspace ,
\ees
(see \fig{f_matrixelements}). 
In the ratio 
\bes
\zp = {\rm const.} \sqrt{f_1} / \fp(x)|_{x_0=L/2}  
\enspace , \label{e_zp}
\ees
the renormalization of the boundary quark fields \myref{StefanI} 
cancels out; it can therefore be taken as the definition of the
renormalization constant of $P$. 
The proportionality 
constant is to be chosen such that $\zp=1$ at tree level. 
Further details of the definition are $C=C'=0$ and a specific choice
for $\theta$. The renormalization constant is to be evaluated
for zero quark mass, $m$. This defines a mass-independent scheme.
%% all 
%%quark masses have the same scale dependence.

By construction, the SF scheme is non-perturbative
and independent of a specific regularization. For a concrete non-perturbative
computation, we do, however, need to evaluate the expectation values
by a MC-simulation of the corresponding lattice theory. For the details of
the lattice formulation we refer to \myref{paper1} but mention that it 
is
essential to use an $\Oa$-improved formulation in order to keep lattice
spacing effects small \myref{Sommer97}. We now explain the computation of 
the scale dependence, omitting the matching to the HS scheme 
\cite{alphaIII,lat97} for lack of space.

\section{Scale evolution, $\Lambda$ and RGI masses}
%renormalization group invariant masses}
%
Each step in the recursive computation
of the scale evolution consists of:\vspace{-0.1cm}
\begin{itemize}
 \item[1.] Choose a lattice with $L/a$ points in each direction.
 \item[2.] Tune the bare coupling, $g_0$, such that the renormalized
           coupling $\gbar^2(L)$ has the value $u$ and tune the 
           bare mass, $m_0$, such that the mass (\eq{e_PCAC}) vanishes;
           compute $\zp$.
 \item[3.] At the same values of $g_0, m_0$, simulate a lattice 
           with twice the linear size; compute 
           $u'=\gbar^2(2L)$ and $\zp'$. This determines the lattice step scaling 
           functions for the coupling, 
           $\Sigma(u,a/L)$=u', and for the mass,
           $\Sigma_{\rm P}(u,a/L)=\zp/\zp'$.
 \item[4.] Repeat steps 1.--3. with different resolutions $L/a$ and 
           extrapolate $\sigma(u)=\lim_{a/L\to0}\Sigma(u,a/L)$,
           $\sigma_{\rm P}(u)=\lim_{a/L\to0}\Sigma_{\rm P}(u,a/L)$.
           This extrapolation can be done by allowing 
           for small linear $a/L$-discretization errors \cite{alphaIII}.      
\end{itemize}\vspace{-0.1cm}
%%%%%%%%%%%%%%%%%%%%%%%%%%%%%FIGURE%%%%%%%%%%%%%%%%%%%%%%%%%%%%%%%%%%%
%%\begin{figure}
%%\vspace{-2.2cm}
%%
%%\psfig{file=plots/alpha_su3.eps,%
%%width=7cm}
%%\vspace{-2.0cm}
%%\caption{The running coupling in  SU(3) Yang-Mills
%%theory.  Uncertainties are smaller than the size of the symbols.
%%\label{f_running}}
%%\end{figure}
%%%%%%%%%%%%%%%%%%%%%%%%%%%%%%%%%%%%%%%%%%%%%%%%%%%%%%%%%%%%%%%%%%%%%%
%%%%%%%%%%%%%%%%%%%%%%%%%%%%%FIGURE%%%%%%%%%%%%%%%%%%%%%%%%%%%%%%%%%%%
\begin{figure}
\vspace{-8.0cm}

\centerline{
\psfig{file=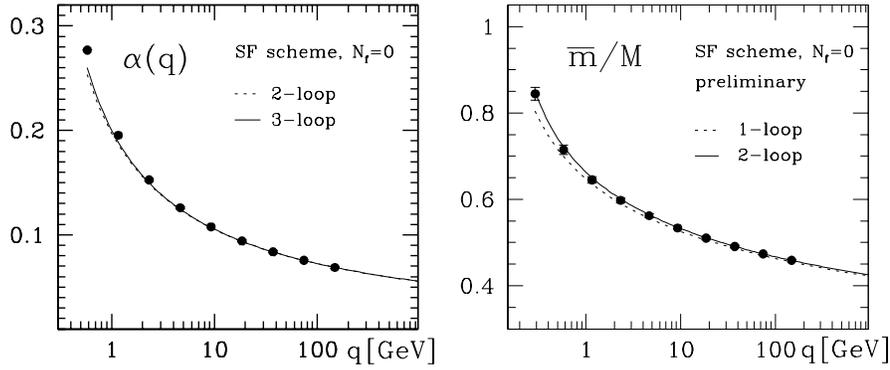,%
width=15.7cm}}
\vspace{-3.3cm}
\caption{Non-perturbative running coupling and quark mass 
as a function of $q\equiv1/L$. 
Uncertainties are smaller than the size of the symbols.
\label{f_running}}
\end{figure}
%%%%%%%%%%%%%%%%%%%%%%%%%%%%%%%%%%%%%%%%%%%%%%%%%%%%%%%%%%%%%%%%%%%%%%
In this way, $\sigma(u), \, 1.1 \leq u \leq 2.4$ and 
$\sigma_{\rm P}(u), \, 1.1 \leq u \leq 3.48$ have been 
computed in the quenched approximation.
The scale evolution is monotonous in this range: $\sigma(u)>u$,
$\sigma_{\rm P}(u) > 1$. $\sigma$ and $\sigma_{\rm P}$ can therefore 
be used to follow \fig{f_strategy}
downwards (up in the energy scale). 
%%Since we are discussing the quenched approximation,
%%the horizontal match to the experimental hadron spectrum is 
%%strictly speaking not 
%%meaningful. Nevertheless, $L_{\rm max}$, the box size
%%with $\gbar^2(\Lmax) = 3.48$,  was related to $r_0$,
%%a length scale defined through the inter-quark force \cite{r0paper}. 
%%For illustration we set $r_0$ to a phenomenological value $r_0=0.5\fm$.
%%Once one
%%tackles full QCD, $r_0$ should be replaced by a direct experimental 
%%observable
%%like $m_{\rm p}$.

Starting from the box size $L_{\rm max}$, defined by $\gbar^2(\Lmax) = 3.48$,
we use
\bes
 \gbar^2(L_{k-1})=\sigma(\gbar^2(L_{k})), \,\,\,
 \mbar(L_{k-1}) = \sigma_{\rm P}(\gbar^2(L_{k}))\, \mbar(L_{k}), 
 \,\,\, L_k=2^{-k}\Lmax
\ees
to
compute the sequences \cite{lat97}
\bes
\gbar^2(L_k),\quad \mbar(L_k)/\mbar(L_{\rm max}), \quad
 k=0,1,\ldots 8 \enspace . 
\ees
In  \fig{f_running}, these are compared to the
perturbative evolutions (starting at the weakest coupling).
It is surprising that the perturbative evolution 
is quite precise down to rather low energy scales. This property may 
of course not be generalized to other schemes but has to be checked
in each scheme separately.

On the other hand, the figure clearly demonstrates that the SF-scheme 
is perturbative in the large energy regime, say $q>50\GeV$. 
We may therefore start from
the exact relations
\bes
 \Lambda &=&q \left(b_0\gbar^2\right)^{-b1/(2b_0^2)} \rme^{-1/(2b_0\gbar^2)}%%\\
         %% && \times
           \exp \left\{-\int_0^{\gbar} \rmd x 
          \left[\frac{1}{ \beta(x)}+\frac{1}{b_0x^3}-\frac{b_1}{b_0^2x}
          \right]
          \right\} \enspace , \label{e_lambdapar} \\
  M &=& \mbar\,(2 b_0\gbar^2)^{-d_0/2b_0^2} 
   \exp \left\{- \int_0^{\gbar} \rmd g [\frac{\tau(g)}{\beta(g)}
     - \frac{d_0}{b_0 g} ] \right\}  \enspace, 
\ees
insert for the  $\beta$-function ($\tau$-function) its 3-loop (2-loop)
approximation and set $q$ to the maximal value reached,
to compute $\Lambda$ and $\mbar/M$. One can easily convince
oneself that the left over perturbative uncertainty
is negligible compared to the accumulated statistical errors.
 
After converting to the $\MSbar$-scheme one arrives at the 
result~\myref{lat97}
\bes
 \Lambda_{\MSbar}^{(0)} = 251 \pm 21 \MeV \enspace, 
 \label{e_lambdares}
\ees
where the label $^{(0)}$ reminds us that this number 
was obtained with zero quark flavors. 
The error in \eq{e_lambdares} accounts for all 
uncertainties including the extrapolations to the continuum limit 
that were done in the intermediate steps.
%%%%%%%%%%%%%%%%%%%%%%%%%%%%%FIGURE%%%%%%%%%%%%%%%%%%%%%%%%%%%%%%%%%%%
%\begin{figure}[ht]
%\vspace{ -2.0cm}
%
%\centerline{
%\psfig{file=plots/mbar.eps,%
%width=7cm}}
%\vspace{-2.0cm}
%\caption{The running quark mass as a function of $q\equiv1/L$.
%\label{f_mbar}}
%\end{figure}
%%%%%%%%%%%%%%%%%%%%%%%%%%%%%%%%%%%%%%%%%%%%%%%%%%%%%%%%%%%%%%%%%%%%
 
This talk is based on research done
in the framework of the ALPHA collaboration\cite{lat97}. I thank in particular
S. Capitani, M. Guagnelli, M. L\"uscher, S. Sint, P. Weisz and H. Wittig
for the joint work.

% ---- Bibliography ----
%

\end{document}